\documentclass[reprint,prl,aps,superscriptaddress,floatfix,showpacs,showkeys,altaffilletter]{revtex4-1}

\newcommand{\nuc}[2]{$^{#1}$#2}

\usepackage{comment}
\usepackage[pdftex]{graphics}      
\usepackage{graphicx}                  
\usepackage[rightcaption]{sidecap}
\usepackage{epsfig}
\usepackage{bm}
\usepackage{verbatim}
\usepackage{color}
\usepackage[percent]{overpic}
\usepackage{multirow}
\usepackage{longtable}
\usepackage{tablefootnote}
\usepackage{footnote}
\usepackage{dcolumn}
\usepackage{setspace}
\usepackage{tabularx}
\usepackage{tablefootnote}
\usepackage{array}
\usepackage{epstopdf}
\usepackage{hyperref}
\usepackage{amsmath}
\newcommand{\RomanNumeralCaps}[1]
    {\MakeUppercase{\romannumeral #1}}
\hypersetup{
     colorlinks   = true,
     citecolor    = blue,
     urlcolor     = blue,
     linkcolor   = blue,
}

\begin{document}


\title{Direct measurement of hexacontatetrapole, \boldmath{$E$}6 \boldmath{$\gamma$} decay from \nuc{\boldsymbol{53m}}{Fe}}


\author{T.~Palazzo}
\affiliation{Department of Nuclear Physics and Accelerator Applications, Research School of Physics, The Australian National University, Canberra, ACT 2601, Australia}

\author{A.~J.~Mitchell}
\email[Email: ]{aj.mitchell@anu.edu.au}
\affiliation{Department of Nuclear Physics and Accelerator Applications, Research School of Physics, The Australian National University, Canberra, ACT 2601, Australia}

\author{G.~J.~Lane}
\affiliation{Department of Nuclear Physics and Accelerator Applications, Research School of Physics, The Australian National University, Canberra, ACT 2601, Australia}

\author{A.~E.~Stuchbery}
\affiliation{Department of Nuclear Physics and Accelerator Applications, Research School of Physics, The Australian National University, Canberra, ACT 2601, Australia}

\author{B.~A.~Brown}
\affiliation{Department of Physics and Astronomy, and the Facility for Rare Isotope Beams, Michigan State University, East Lansing, Michigan 48824-1321, USA}

\author{M.~W.~Reed}
\affiliation{Department of Nuclear Physics and Accelerator Applications, Research School of Physics, The Australian National University, Canberra, ACT 2601, Australia}

\author{A.~Akber}
\affiliation{Department of Nuclear Physics and Accelerator Applications, Research School of Physics, The Australian National University, Canberra, ACT 2601, Australia}

\author{B.~J.~Coombes}
\affiliation{Department of Nuclear Physics and Accelerator Applications, Research School of Physics, The Australian National University, Canberra, ACT 2601, Australia}

\author{J.~T.~H.~Dowie}
\affiliation{Department of Nuclear Physics and Accelerator Applications, Research School of Physics, The Australian National University, Canberra, ACT 2601, Australia}

\author{T.~K.~Eriksen}
\affiliation{Department of Nuclear Physics and Accelerator Applications, Research School of Physics, The Australian National University, Canberra, ACT 2601, Australia}

\author{M.~S.~M.~Gerathy}
\affiliation{Department of Nuclear Physics and Accelerator Applications, Research School of Physics, The Australian National University, Canberra, ACT 2601, Australia}

\author{T.~Kib\'{e}di}
\affiliation{Department of Nuclear Physics and Accelerator Applications, Research School of Physics, The Australian National University, Canberra, ACT 2601, Australia}

\author{T.~Tornyi}
\affiliation{Department of Nuclear Physics and Accelerator Applications, Research School of Physics, The Australian National University, Canberra, ACT 2601, Australia}

\author{M.~O.~de~Vries}
\affiliation{Department of Nuclear Physics and Accelerator Applications, Research School of Physics, The Australian National University, Canberra, ACT 2601, Australia}

\date{\today}


\begin{abstract}

The only proposed observation of a discrete, hexacontatetrapole ($E6$) transition in nature occurs from the T$_{1/2}$~=~2.54(2)-minute decay of \nuc{53m}{Fe}. However, there are conflicting claims concerning its $\gamma$-decay branching ratio, and a rigorous interrogation of $\gamma$-ray sum contributions is lacking. Experiments performed at the Australian Heavy Ion Accelerator Facility were used to study the decay of \nuc{53m}{Fe}. For the first time, sum-coincidence contributions to the weak $E6$ and $M5$ decay branches have been firmly quantified using complementary experimental and computational methods. Agreement across the different approaches confirms the existence of the real $E6$ transition; the $M5$ branching ratio and transition rate have also been revised. Shell model calculations performed in the full $pf$ model space suggest that the effective proton charge for high-multipole, $E4$ and $E6$, transitions is quenched to approximately two-thirds of the collective $E2$ value. Correlations between nucleons may offer an explanation of this unexpected phenomenon, which is in stark contrast to the collective nature of lower-multipole, electric transitions observed in atomic nuclei.

\end{abstract}


\maketitle



First-order electromagnetic processes are the primary mechanism by which excited states in atomic nuclei relax, most often via single $\gamma$-ray emission. Since both initial- and final-state wave functions possess a well-defined spin ($J$) and parity ($\pi$), conservation laws impose a characteristic multipolarity ($\sigma \lambda$) for each discrete transition. Nature favours pathways that proceed via the lowest available multipole order; as such, $\Delta J = 1, 2$ transitions are prevalent in atomic and nuclear systems. However, situations arise in which the only available decay pathway is hindered by a larger angular-momentum-change requirement \cite{walker1999}. As the multipole order increases, the number of known cases decreases rapidly. For example, there are $\approx$ 1100 pure or mixed $\Delta J=3$ ($E3$ or $M3$), $\approx$ 170 $\Delta J=4$ ($E4$ or $M4$), and $\approx$ 25 $\Delta J=5$ ($E5$ or $M5$) transitions reported in atomic nuclei.

\begin{figure*}[t!]
\includegraphics[width=17.5 cm]{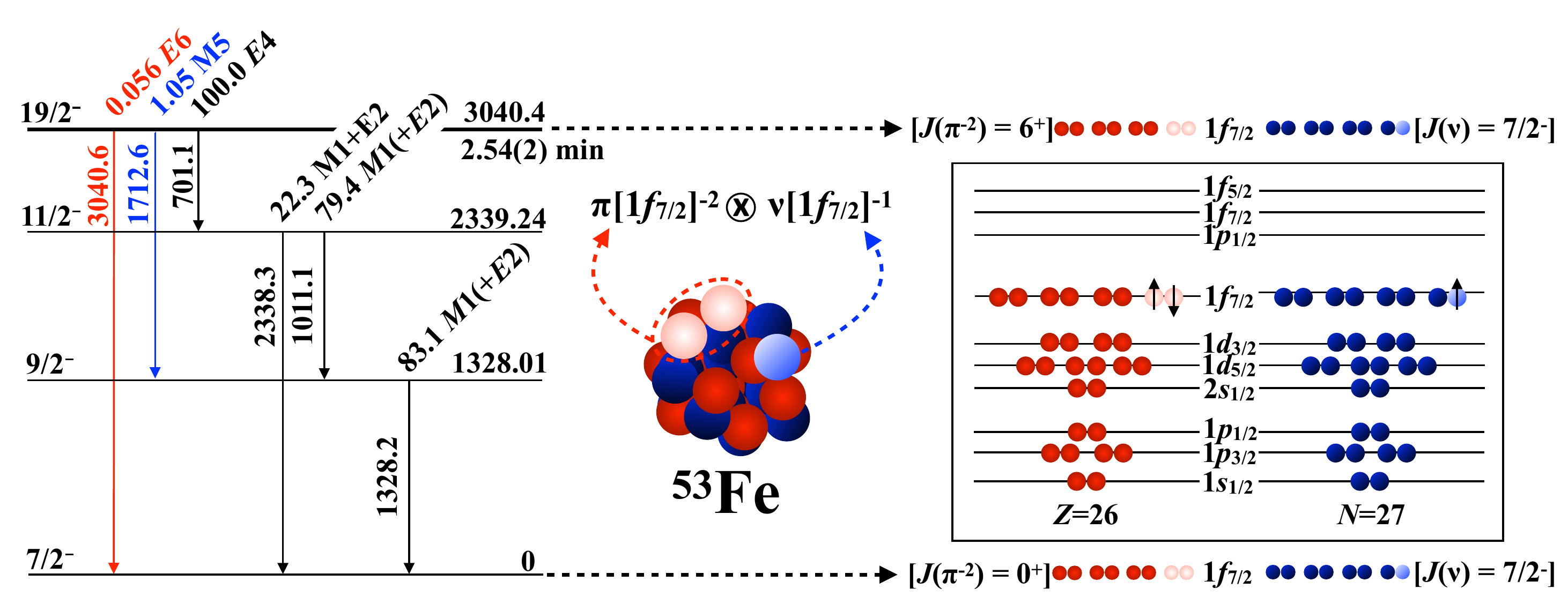}
\caption{\label{fig1} Level scheme showing the energies (in keV) of excited states and $\gamma$-ray transitions observed in the decay of \nuc{53m}{Fe} \cite{junde}, together with nucleon configurations that couple to form the 19/2$^-$ isomer. The $\gamma$-ray intensities were determined in this work. Proton (neutron) particles are depicted by red (blue) solid spheres; proton (neutron) holes are shown as faded spheres. Coupling of the proton- and neutron-hole configurations  leads to formation of the 19/2$^-$ isomeric state at 3040~keV.}
\end{figure*}

Despite discovery of over 3,000 different nuclides, only one claim of $\Delta J=6$, or hexacontatetrapole, decay has been reported: the $J^{\pi} = 19/2^- \to J^{\pi} = 7/2^-$, $E6$ $\gamma$ decay from \nuc{53m}{Fe} \cite{black1971,black1975,geesaman1976,geesaman1979} (see Fig.~\ref{fig1} for details). Low-lying states in this nucleus can be understood in the $ (f_{7/2})$ model space with an effective interaction derived from the energy-level spectra of $^{54}$Co (\nuc{53}{Fe} plus a proton) and $^{54}$Fe (\nuc{53}{Fe} plus a neutron) \cite{geesaman1976}. Isomerism of the 19/2$^{-}$ level occurs due to its location relative to the other yrast states i.e., those with the lowest excitation energy for a given spin and parity. The only alternate decay pathways to the $E6$ transition are the strongly hindered $M5$, $J^{\pi} = 19/2^- \to 9/2^-$ and $E4$, $J^{\pi} = 19/2^- \to 11/2^-$ transitions.

However, inconsistencies in $\gamma$-ray branching ratios and reduced transition rates are reported in the literature \cite{black1971,black1975}. Although they are relatively rare, $\gamma$-ray `summing' events could be mistaken for the very weak, $E6$ decay; these occur when multiple $\gamma$ rays are incident on the same detector within an unresolvable time window. It is even possible that no real $E6$ transition was observed in the prior work, and the feature at 3041~keV reported in the energy spectrum of Ref.~\cite{black1971} consists entirely of sum events. Despite their importance, a thorough and quantitative understanding of sum contributions was lacking \cite{black1971,black1975}.

This Letter reports the first direct confirmation of $E6$ $\gamma$ decay in \nuc{53m}{Fe} using a novel combination of experimental, computational and Monte Carlo techniques that fully quantify the sum contributions; this confirms the highest multipole order ever observed. With a now-well-defined $E6$ transition strength, and revised values for the $M5$ and $E4$ $\gamma$ decay, \nuc{53m}{Fe} provides a unique test of the nuclear shell model and our present understanding of high-multipolarity transitions within a single nuclear system. Comparison with theoretical shell model calculations performed in the full $fp$-model space shows, surprisingly, that low- and high-multipolarity transitions in atomic nuclei are fundamentally different in nature. 


The experiments were performed at the Heavy Ion Accelerator Facility at the Australian National University. A 2-pnA beam of 50-MeV, \nuc{6}{Li} ions delivered by the 14UD Pelletron accelerator was incident on self-supporting targets of natural vanadium. Three separate, 10-mg/cm$^2$ thick targets were used; these were replaced periodically to suppress build up of long-lived activity. Excited states in \nuc{53}{Fe} were populated via the \nuc{51}{V}(\nuc{6}{Li},4$n$)\nuc{53}{Fe} reaction. Other fusion-evaporation channels led to production of neighbouring isotopes of iron, manganese, chromium, vanadium, titanium and scandium. Since many of these nuclides are stable against $\beta$~decay, their prompt $\gamma$ rays were easily separated from delayed decay of \nuc{53m}{Fe} via subtraction of suitable sections of the time-correlated data discussed below. 

Relaxation of \nuc{53m}{Fe} was studied via $\gamma$-ray spectroscopy using the CAESAR array of Compton-suppressed High-Purity Germanium (HPGe) detectors \cite{caesar}. Of the nine detectors used, six were fixed in the vertical plane, perpendicular to the beam axis and $\approx$ 12~cm from the target. The remaining three, in the horizontal plane, were on rail systems allowing their radial position to be moved. The detector-suppressor assemblies were retracted such that the front collimator that defines the detector illumination was moved from $\approx$~8.5~cm to $\approx$~12~cm from the target between measurements, reducing the exposed solid angle by approximately a factor of two. These are referred to as the `near' and `far' geometries, respectively, and discussed quantitatively in the text below. Standard $\gamma$-ray sources of \nuc{152}{Eu} and \nuc{56}{Co} were used for energy and absolute detection-efficiency calibrations.

A continuous \nuc{6}{Li} beam irradiated the target for 7.5~minutes (approximately three half-lives of \nuc{53m}{Fe}), after which the beam was intercepted and decay of the isomer was observed for 20~minutes (approximately eight half-lives). A custom-made counter, with an oscillator that can be driven at various well-defined frequencies, was used in conjunction with the CAESAR data acquisition system to time-stamp individual $\gamma$-decay events across many repeating irradiation-decay cycles. Observation of intense 701-, 1011-, 1328- and 2338-keV $\gamma$ rays confirmed production of \nuc{53m}{Fe}.

\begin{figure}[t!]
\includegraphics[width=8.5 cm]{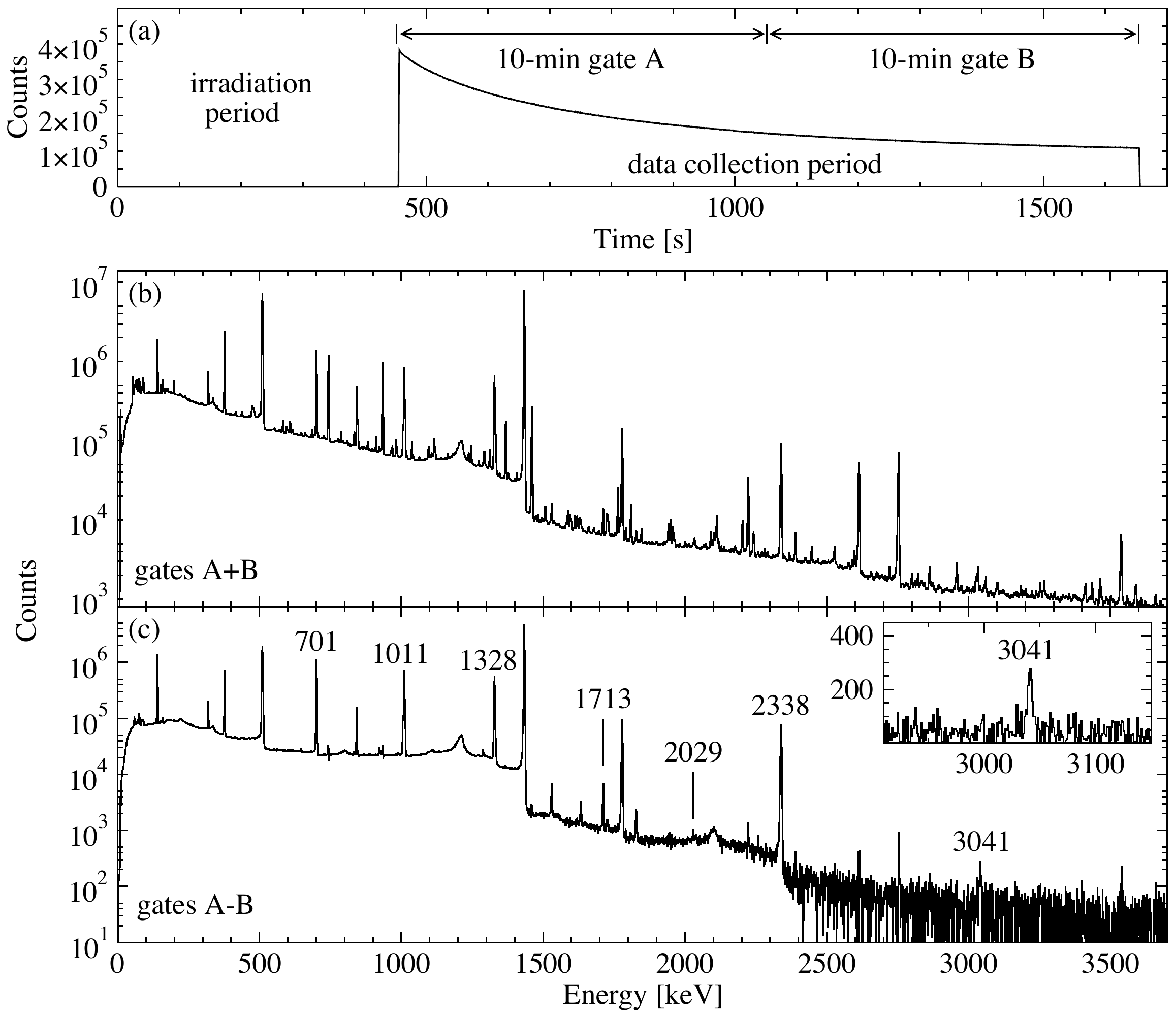}
\caption{\label{fig2} (a) Time spectrum from the ADC clock recorded with each $\gamma$-ray event illustrating the irradiation and out-of-beam collection period split into two parts, gates A and B. Lower panels show (b) the total $\gamma$-ray spectrum recorded (gate A plus gate B) and (c) the subtracted spectrum (gate A minus gate B) described in the text. The inset spectrum is on a linear scale and expands the region near the 3041-keV, $E6$ transition.}
\end{figure}

The bulk of nuclei produced in the reactions have much longer lifetimes than \nuc{53m}{Fe}. Subtracting the second 10 minutes of the collection cycle from the first 10 minutes resulted in a much cleaner energy spectrum that strongly enhances the peak-to-total ratio for \nuc{53m}{Fe} decay, while only sacrificing $\approx$~12$\%$ of the total \nuc{53m}{Fe} data collected. The time spectrum of collected events, as well as the total $\gamma$-ray and time-subtracted $\gamma$-ray energy spectra are presented in Fig.~\ref{fig2}. Gamma rays from the decay of \nuc{53m}{Fe} have been labeled by their energy in keV. The remaining $\gamma$~rays have been identified as arising from decay of \nuc{75m}{Ge}~(T$_{1/2}~=~48$~s), and $\beta$~decay of \nuc{51}{Ti}~(T$_{1/2}~=~346$~s), \nuc{53}{Fe}~(ground state, T$_{1/2}~=~510$~s), \nuc{52}{V}~(T$_{1/2}~=~208$~s), \nuc{20}{F}~(T$_{1/2}~=~11$~s) and \nuc{28}{Al}~(T$_{1/2}~=~134$~s).

Total yields of $\gamma$ rays from \nuc{53m}{Fe} decay, measured in both geometries, are provided in Table~I of Ref.~\cite{supp}. In addition to the real $E6$ transition reported in this Letter, \nuc{53m}{Fe} exhibits three alternate decay pathways to the ground state (refer to Fig.~\ref{fig1} for details). Each individual cascade presents a potential summing contribution ($S_{i}^{}$) to the true 3041-keV $\gamma$-ray intensity ($I_{\gamma}^{}$) that requires careful consideration. The observed full-energy peak yield $(Y_{\gamma}^{})$ is given by: \vspace{-0.4cm}

\begin{equation}
Y_{\gamma}^{}~=~I_{\gamma}^{}~+~\Sigma S_{i}^{},  \label{equation1} \\
\end{equation}

\noindent
where the sum is over each possible multi-transition cascade that connects the level to the ground state. While the real 1713-, 2338- and 3041-keV full-energy peaks are all expected to contain individual sum contributions, an additional peak observed at 2029 keV in Fig.~\ref{fig2} is entirely composed of sum events (701~keV + 1328~keV).

Experimental and computational methods were adopted to quantify the sum-coincidence component in each of these measured full-energy peak yields. Full details of the methods and their results are described in Refs.~\cite{palazzo, supp}; a brief explanation of each method is provided here:

\noindent
$\bullet$ $Experimental$: The measured yield of the 2029-keV full-energy sum peak, which can $only$ occur though summing, can be scaled to estimate the sum-coincidence components of the other transitions while accounting for detection efficiencies and angular correlations. \\
$\bullet$  $Geometric$: Sum-coincidence events can be directly inferred by considering changes in counting efficiency between the `near' and `far' detector geometries. \\
$\bullet$  $Computational$: The sum contribution to $Y_{\gamma}$(3041 keV) can be estimated from measured $\gamma$-ray intensities, detection efficiencies and angular correlations by solving the set of equations that govern the different sum contributions. \\
$\bullet$  $Monte~Carlo$: A Monte Carlo simulation was developed to model the $\gamma$ decay of \nuc{53m}{Fe} and evaluate summing contributions expected with the CAESAR array. 

Consistency between the various approaches across both detector geometries gives confidence in the deduced branching ratios. Therefore, the analysis confirms that the $E6$ transition is real, and enables a firm measurement of its decay branching ratio for the first time.

\begin{table*}[t!]
\centering
\caption{Summary of adopted level and $\gamma$-ray energies, transition multipolarities, newly measured relative intensities (taking sum-coincidence events into account) and deduced transition strengths for the $E4$, $M5$, and $E6$ measured in this work quoted in units of Weisskopf units (W.u), as well as e$^2$fm$^{2\lambda}$ for the $E4$ and $E6$ transitions and $\mu^2_N$ fm$^{2\lambda -2}$ for the $M5$. The half-life of the $J^{\pi}$~=~19/2$^-$ isomer is 2.54(2)~minutes \cite{junde}. Conflicting relative intensities quoted in Table 1 of Ref.~\cite{black1971} and Table III of Ref.~\cite{black1975} are provided for reference. Transition strengths calculated using the branching ratios of Ref.~\cite{black1975} are also provided for comparison with those of the present work.}
\vspace{0.1cm}
\label{table1}
\begin{tabular*}{18cm}{@{\extracolsep{\fill}}cccrrrrrrr}
\hline
\hline \\[-0.3cm]
$E_{\rm{Level}}$	&	$E_\gamma$	& $\sigma L$	&	\multicolumn{3}{c}{$I_\gamma$}	&  \multicolumn{2}{c}{$B(\sigma\lambda)$ (W.u)}  &  \multicolumn{2}{c}{$B(\sigma\lambda)$ (e$^2$fm$^{2\lambda}$, $\mu^2_N$ fm$^{2\lambda -2}$)}  \\[0.1cm]
\hline \\[-0.3cm]
Ref.~\cite{junde}& Ref.~\cite{junde} &	Ref.~\cite{junde}&	This work	 	& Ref.~\cite{black1971} &	Ref.~\cite{black1975} 	&	This work	   &   $I_\gamma$(\cite{black1975})	&	This work	 &	$I_\gamma$(\cite{black1975}) 	 \\[0.1cm]
\hline
\hline \\[-0.3cm]

3040.4 	&	  701.1(1)		& $E4$		&	$\equiv$100	&	$\equiv$100	&		$\equiv$100		& 0.2593(21) 	& 0.2587(21)  	& 6.46(5)$\times$10$^2$ 		& 6.44(6)$\times$10$^2$ 	\\
 		&	  1712.6(3)	& $M5$		&		1.05(5)  	&	0.7(1)		&		1.3(1)			& 4.34(21)  	& 5.4(4)   		& 3.31(16)$\times$10$^5$	& 4.1(3)$\times$10$^5$    	\\
 		&	  3040.6(5)	& $E6$		&		0.056(17) 	&	0.020(5)		&		0.06(1)			& 0.42(12)     	& 0.45(8)   	& 2.61(81)$\times$10$^5$	& 2.8(5)$\times$10$^5$     		\\
2339.24 	&	  1011.2(2)	& $M$1(+$E$2)&		79.4(3) 	&	86(9)			&	86(9)			& &	& & \\
 		&	  2338.3(5)	& $M$1+$E$2	&		22.3(2)	 &	13(2)			&	13(2)			& & & & 	\\[0.1cm]

\hline
\hline
\end{tabular*}
\end{table*}

Transition strengths for the $E4$, $M5$ and $E6$ decays were calculated using the new branching ratios derived from results of the Experimental method; they are presented in Table~\ref{table1}. These have been determined using the adopted 19/2$^-$ state lifetime of T$_{1/2}$~=~2.54(2)~min \cite{junde} and theoretical internal conversion coefficients; values for $L = 1-5$ were calculated using \textsc{BRICC} \cite{kibedi}, while for $L=6$ it was calculated directly using the \textsc{RAINE} code \cite{band2002}. Intensities reported by Black $et~al$ \cite{black1971, black1975}, and transition strengths determined using the relative intensities of Ref.~\cite{black1975} are included for comparison. We confirm the reported values for $E4$ decay, however, the competing $M5$ branching ratio and transition strength were found to be $\approx$20$\%$ lower. Notably, the branching ratios of transitions depopulating the state at 2339~keV were also found to be significantly different to those of Black~$et~al$ \cite{black1975}.\\


To gain microscopic understanding of the high-multipolarity transitions in $^{53m}$Fe, shell model calculations were performed with the \textsc{NuShellX} code \cite{brown2014}. For comparisons between theory and experiment, it is useful to consider the reduced matrix element, $\mathcal{M}_p$, which is related to the reduced transition strength by: \\[-0.8cm]

\begin{equation}
B(E\lambda; J_i \to J_f) = \frac{\mathcal{M}_p^2}{(2J_i~+~1)},
\end{equation}

\noindent
where $\mathcal{M}_p$ is further separated into its proton ($\mathcal{A}_p$) and neutron ($\mathcal{A}_n$) contributions: \\[-0.8cm]

\begin{equation}
\mathcal{M}_p~=~\mathcal{A}_p \cdot \varepsilon_p~+~\mathcal{A}_n \cdot \varepsilon_n.
\label{nme}
\end{equation}

Typically, $\mathcal{A}_p$ and $\mathcal{A}_n$ are calculated to account for configuration mixing within the major shell, while effective nucleon charges are introduced to account for cross-shell mixing. Thus $\varepsilon_{p,n} = e_{p,n} + \delta_{p,n}$, where $e_{p,n}$ are bare nucleon charges and $\delta_{p,n}$ are core-polarization charges.


Calculations were performed within a restricted ($f_{7/2})^{13}$, and full $ fp $ model space with two commonly used Hamiltonians, GFPX1A \cite{honma2005} and KB3G \cite{poves2001}. Excited-state energies were in good agreement with the adopted values \cite{junde}; for example, the energies of the $19/2^-$, $11/2^-$ and $9/2^-$ states calculated with the GFPX1A interaction have a root-mean-squared (rms) deviation of 169 keV. Matrix elements for the electromagnetic transitions are sensitive to the rms radius of the $ 0f_{7/2} $ orbit, and with harmonic oscillator radial wavefunctions they scale approximately with $b^{\lambda}$, where $b$ is the oscillator length parameter. Spherical Skyrme Hartree-Fock calculations, with Skx \cite{brown1998} and Sly4 \cite{brown1983} interactions, were used to determine the $ 0f_{7/2} $ orbital rms radius. The Skx $ 0f_{7/2} $ rms radius was reproduced by the harmonic oscillator model with $  b=1.937 $~fm. This parameter is approximately 3\% larger for Sly4, which represents the theoretical uncertainty in the rms radius. The matrix elements, therefore, have uncertainties of 18\%, 15\%, and 12\% for the calculated $ \lambda=6,5, 4 $ matrix elements, respectively.

The full set of results is provided in Table~\RomanNumeralCaps{2} of Ref.~\cite{supp}, and average values of both $ fp $-shell calculations are summarised and compared to experiment in Table~\ref{table2} in this paper. Results of the ($f_{7/2})^{13}$ calculations are similar to those in Ref.~\cite{gloeckner1975}. Surprisingly, matrix elements obtained in the full $ fp $ model space are almost a factor of two smaller than the restricted-basis values. This is unusual, since strong $ \lambda =2 $ transitions are generally enhanced in the full $ fp $ space with respect to the restricted one. This behavior comes about because the high-$\lambda$ transitions are dominated by the $ 0f_{7/2} $ orbital; in the larger space, the matrix elements are diluted by mixing of the $ 0f_{7/2} $ component with 1$p$ orbitals, which cannot contribute to the high-multipolarity transitions; in contrast, the $ 1p $ orbitals contribute to and enhance $ \lambda=2 $ transition strength.

\begin{table}[t]
\begin{center}
\caption{\label{table2} Theoretical values of proton and neutron contributions to the $E4$, $M5$ and $E6$ matrix elements ($\mathcal{A}_{p,n}$) calculated in the full $fp$ model space, discussed in the text. Uncertainties in the calculated matrix elements are $\pm$(18,15,12)\% for $ \lambda=(6,5,4) $, respectively. For the $M5$ transition, $\mathcal{M}~=~(\mathcal{A}_p~+~\mathcal{A}_n)$. Experimental matrix elements ($\mathcal{M}_p^{\rm{expt.}}$) are determined from this work.}
\renewcommand{\arraystretch}{1.5}
\begin{tabular*}{0.48\textwidth}{@{\extracolsep{\fill} }crrrr}
\toprule
$\sigma L$ 	&	$\mathcal{A}_p\times10^3$	&	$\mathcal{A}_n\times10^3 $	&	$\mathcal{M}\times10^3$	&	$\mathcal{M}_p^{\rm{expt.}}\times10^3$	  \\[0.05cm]
\hline
$E4$		& 	0.142(17)		&	0.045(7)		& 	-	&	0.1137(5)			   \\
$M5$	& 	5.09(76)	&	-0.11(2)		& 	4.98(76)			&	2.57(6) 	  		 \\
$E6$		& 	3.52(63)	&	0.22(4)	& 	-	&	2.29(35)		  		  \\
\hline
\hline
\end{tabular*}
\end{center}
\end{table}

A remarkable aspect of these high-multipolarity transitions is that they are dominated by their proton component. This, again, is in contrast to strong $ B(E2) $ transitions, in which the proton and neutron components are typically observed to be similar. For this reason, the isoscalar $ E2 $ effective charge is best determined with, for example, the empirical value of $ \varepsilon_{p}+\varepsilon_{n}=2.0 $ obtained in Ref.~\cite{honma2004}. The separate proton and neutron $ E2 $ effective charges can only be obtained in special cases. An example is the $ A~=~51 $ mirror nuclei system \cite{durietz2004}, where values of $ \varepsilon_{p} \approx 1.15 $ and $ \varepsilon_{n}\approx 0.80 $ were obtained from the measured $ E2 $ transition data.

\begin{figure}[t!]
\includegraphics[width=8.5 cm]{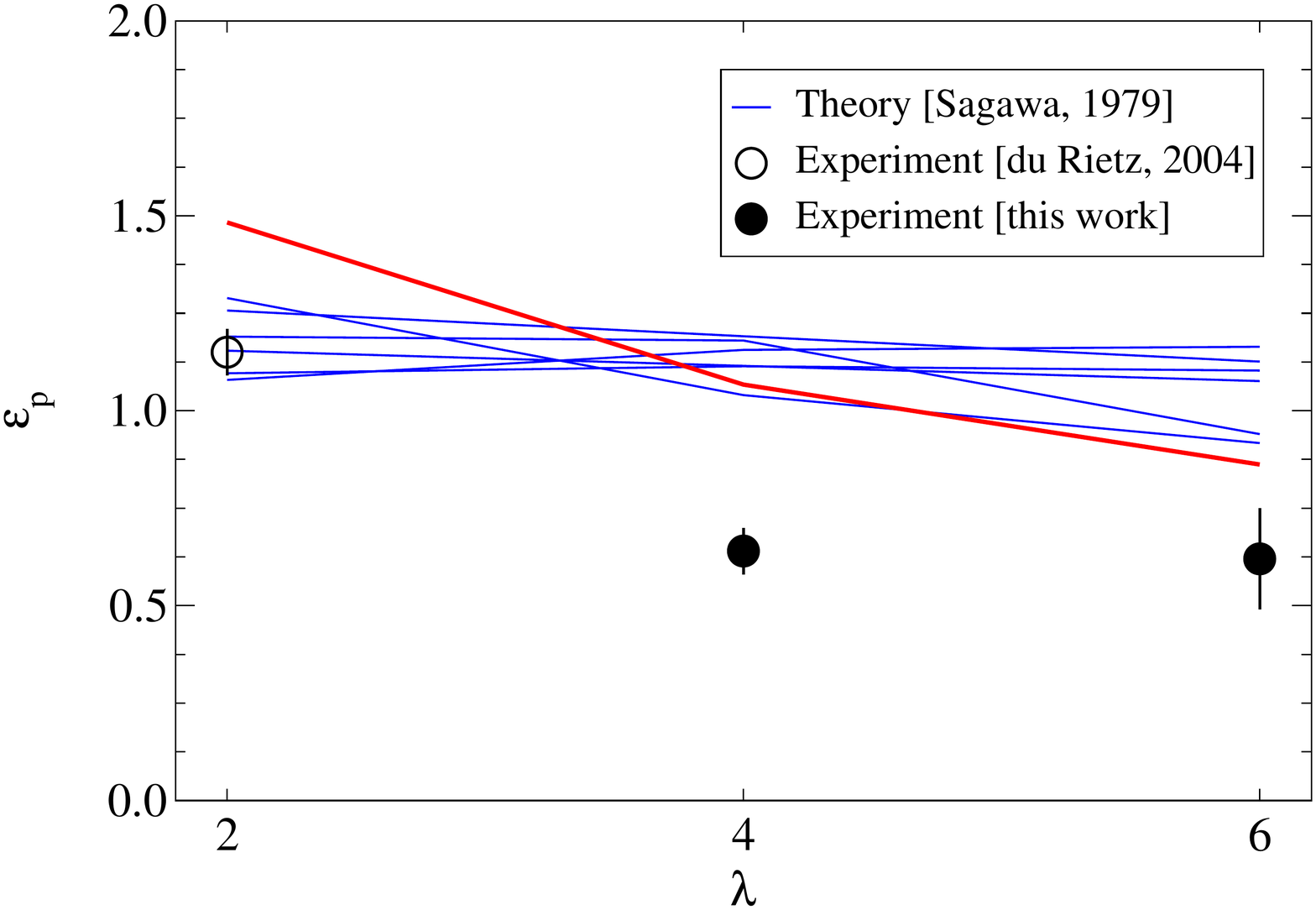}
\caption{\label{fig3} Proton effective charges calculated for $ \lambda = 2, 4, \textrm{and}~6$ with seven different interactions (red and blue lines) \cite{sagawa1979} compared to experimental values for $ \lambda = 2 $ (open circle) \cite{durietz2004} and $ \lambda = 4, 6$ (closed circles) from this work.}
\end{figure}

The calculated proton and neutron contributions and experimental matrix elements, presented in Table~\ref{table2}, can be used with Equation~\eqref{nme} to obtain effective proton charges for the high-multipolarity electric transitions. For the small neutron component, $ \varepsilon_{n}=0.5 $ is adopted \cite{sagawa1979}. The results obtained are: $ \varepsilon_{p}=0.62(13) $ for $ \lambda=6 $; and $ \varepsilon_{p}=0.64(6) $ for $ \lambda=4 $; if a value of $ \varepsilon_{n}=0 $ is used instead, $ \varepsilon_{p} = 0.65(13) $ and $ \varepsilon_{p} = 0.80(7) $ are found for $ \lambda=6 $ and $ \lambda=4 $, respectively. These results are presented in Fig.~\ref{fig3}, along with the value of $ \varepsilon_{p}=1.15 $ for $ \lambda =2 $ from Ref.~\cite{durietz2004}, which has an assumed uncertainty of 5\%.

Effective charges are evaluated by considering the coupling of valence nucleons to particle-hole excitations of the core. Whether based on perturbation theory or the particle-vibration concepts of Bohr and Mottelson \cite{BM}, there is a choice of---and sensitivity to---the residual particle-hole interaction adopted in the calculation. Core-polarization contributions for all $ \lambda $ values were calculated for seven different interactions in Ref.~\cite{sagawa1979}. The results of these calculations, summarized in Table~\RomanNumeralCaps{1} of Ref.~\cite{sagawa1979}, are compared to empirical values for $\lambda = 2, 4, 6$ in Fig.~\ref{fig3}. The one that adopts Wigner-type interactions, shown in red, has a trend which is closest matched to experiment. However, while there is excellent agreement for $ \lambda=2 $, all of the theoretical results are too large for $ \lambda=4 $ and $ \lambda=6 $.

The $E6$ matrix element within the ($0f_{7/2})^{13}$ configuration can be written as a product of two 0$f_{7/2}$ spectroscopic amplitudes for one-proton removal times the single-particle $E6$ matrix element. Cross sections from ($e,e'p$) data are also proportional to the product of two 0$f_{7/2}$ spectroscopic amplitudes; these are quenched by about a factor of two compared to those calculated in the $fp$ model space (see e.g., Ref.~\cite{denherder1986} for $^{51}$V($e,e'p$)$^{50}$Ti). This is interpreted as a ``dilution'' of the $fp$ part of the wavefunction due to short- \cite{muther1995, dickhoff2010} and long-range \cite{barbieri2009} correlations that go beyond the $fp$ model-space. This phenomenon is observed more broadly across the nuclear landscape \cite{lapikas1993, kramer2001} and cross sections extracted from nucleon transfer-reaction data are also known to be quenched by a similar magnitude \cite{kay2013}. The similarities suggest that quenching of the $E6$ matrix element observed in this work and quenching of ($e,e'p$) cross sections are connected. Ultimately, any model that is used to understand the quenching of nucleon-removal cross sections should be extended to include calculations of electromagnetic matrix elements.

Since matrix elements of single-particle operators can be expanded in terms of the overlap integrals between eigenstates of a system with $ A $ nucleons and one of mass $ (A-1) $ \cite{berggren1965}, high-multipole transitions appear to provide a sensitive probe of single-particle features of atomic nuclei. Further theoretical investigation into the high-multipolarity matrix elements, that includes such correlations, is therefore necessary.

In summary, experimental observation of an $E6$ transition in $^{53}$Fe is unambiguously confirmed by identifying and removing sum-coincidence contributions with three distinct methods that are in mutual agreement. Transition strengths for the high-multipolarity transitions from the 2.54(2)-minute, $J$~=~19/2$^-$ isomer have been determined from the newly measured branching ratios. In the $fp$ model space, the $E6$ strength comes mainly from the dominant $(0f_{7/2})^{13}$ configuration. When this mixes with the many other $fp$ configurations, the $(0f_{7/2})^{13}$ configuration becomes `diluted' and the total $E6$ matrix element decreases by about a factor of two in our calculations. The negative effective charge obtained for the full $fp$ model space for $E6$ could be connected as a further dilution relative to the `exact' wavefunction that goes beyond the $fp$ model space. Connection of the reduction of ($e,e'p$) cross sections compared to those calculated in the $fp$ model space was also discussed.


\begin{acknowledgments}
The authors are grateful for excellent support from technical staff of the Department of Nuclear Physics and Accelerator Applications, ANU and the Australian Heavy Ion Accelerator Facility. We thank J. Heighway for preparing targets for these experiments. This work was supported by the Australian Research Council Grants No.~DP170101673 and No.~DP170101675, the International Technology Center Pacific (ITC-PAC) under Contract No. FA520919PA138, and NSF Grant PHY-2110365. A.A., B.J.C., J.T.S.D., M.S.M.G, and T.P. acknowledge support of the Australian Government Research Training Program. Support for the ANU Heavy Ion Accelerator Facility operations through the Australian National Collaborative Research Infrastructure Strategy program is acknowledged. Figure~\ref{fig1} in this letter was created using the LevelScheme scientific figure preparation system \cite{caprio2005}. \\
\end{acknowledgments}

\linespread{1}


\begin{thebibliography}{30}%
\makeatletter
\providecommand \@ifxundefined [1]{%
 \@ifx{#1\undefined}
}%
\providecommand \@ifnum [1]{%
 \ifnum #1\expandafter \@firstoftwo
 \else \expandafter \@secondoftwo
 \fi
}%
\providecommand \@ifx [1]{%
 \ifx #1\expandafter \@firstoftwo
 \else \expandafter \@secondoftwo
 \fi
}%
\providecommand \natexlab [1]{#1}%
\providecommand \enquote  [1]{``#1''}%
\providecommand \bibnamefont  [1]{#1}%
\providecommand \bibfnamefont [1]{#1}%
\providecommand \citenamefont [1]{#1}%
\providecommand \href@noop [0]{\@secondoftwo}%
\providecommand \href [0]{\begingroup \@sanitize@url \@href}%
\providecommand \@href[1]{\@@startlink{#1}\@@href}%
\providecommand \@@href[1]{\endgroup#1\@@endlink}%
\providecommand \@sanitize@url [0]{\catcode `\\12\catcode `\$12\catcode
  `\&12\catcode `\#12\catcode `\^12\catcode `\_12\catcode `\%12\relax}%
\providecommand \@@startlink[1]{}%
\providecommand \@@endlink[0]{}%
\providecommand \url  [0]{\begingroup\@sanitize@url \@url }%
\providecommand \@url [1]{\endgroup\@href {#1}{\urlprefix }}%
\providecommand \urlprefix  [0]{URL }%
\providecommand \Eprint [0]{\href }%
\providecommand \doibase [0]{https://doi.org/}%
\providecommand \selectlanguage [0]{\@gobble}%
\providecommand \bibinfo  [0]{\@secondoftwo}%
\providecommand \bibfield  [0]{\@secondoftwo}%
\providecommand \translation [1]{[#1]}%
\providecommand \BibitemOpen [0]{}%
\providecommand \bibitemStop [0]{}%
\providecommand \bibitemNoStop [0]{.\EOS\space}%
\providecommand \EOS [0]{\spacefactor3000\relax}%
\providecommand \BibitemShut  [1]{\csname bibitem#1\endcsname}%
\let\auto@bib@innerbib\@empty
\bibitem [{\citenamefont {Walker}\ and\ \citenamefont
  {Dracoulis}(1999)}]{walker1999}%
  \BibitemOpen
  \bibfield  {author} {\bibinfo {author} {\bibfnamefont {P.}~\bibnamefont
  {Walker}}\ and\ \bibinfo {author} {\bibfnamefont {G.~D.}\ \bibnamefont
  {Dracoulis}},\ }\href {https://doi.org/https://doi.org/10.1038/19911}
  {\bibfield  {journal} {\bibinfo  {journal} {Nature}\ }\textbf {\bibinfo
  {volume} {399}},\ \bibinfo {pages} {35} (\bibinfo {year} {1999})}\BibitemShut
  {NoStop}%
\bibitem [{\citenamefont {Black}\ \emph {et~al.}(1971)\citenamefont {Black},
  \citenamefont {McHarris},\ and\ \citenamefont {Kelly}}]{black1971}%
  \BibitemOpen
  \bibfield  {author} {\bibinfo {author} {\bibfnamefont {J.~N.}\ \bibnamefont
  {Black}}, \bibinfo {author} {\bibfnamefont {W.~C.}\ \bibnamefont
  {McHarris}},\ and\ \bibinfo {author} {\bibfnamefont {W.~H.}\ \bibnamefont
  {Kelly}},\ }\href {https://doi.org/10.1103/PhysRevLett.26.451} {\bibfield
  {journal} {\bibinfo  {journal} {Phys. Rev. Lett.}\ }\textbf {\bibinfo
  {volume} {26}},\ \bibinfo {pages} {451} (\bibinfo {year} {1971})}\BibitemShut
  {NoStop}%
\bibitem [{\citenamefont {Black}\ \emph {et~al.}(1975)\citenamefont {Black},
  \citenamefont {McHarris}, \citenamefont {Kelly},\ and\ \citenamefont
  {Wildenthal}}]{black1975}%
  \BibitemOpen
  \bibfield  {author} {\bibinfo {author} {\bibfnamefont {J.~N.}\ \bibnamefont
  {Black}}, \bibinfo {author} {\bibfnamefont {W.~C.}\ \bibnamefont {McHarris}},
  \bibinfo {author} {\bibfnamefont {W.~H.}\ \bibnamefont {Kelly}},\ and\
  \bibinfo {author} {\bibfnamefont {B.~H.}\ \bibnamefont {Wildenthal}},\ }\href
  {https://doi.org/10.1103/PhysRevC.11.939} {\bibfield  {journal} {\bibinfo
  {journal} {Phys. Rev. C}\ }\textbf {\bibinfo {volume} {11}},\ \bibinfo
  {pages} {939} (\bibinfo {year} {1975})}\BibitemShut {NoStop}%
\bibitem [{\citenamefont {Geesaman}(1976)}]{geesaman1976}%
  \BibitemOpen
  \bibfield  {author} {\bibinfo {author} {\bibfnamefont {D.}~\bibnamefont
  {Geesaman}},\ }\emph {\bibinfo {title} {Spin gap isomers in $^{52}$Fe,
  $^{53}$Fe, and $^{54}$Co}},\ \href@noop {} {Ph.D. thesis},\ \bibinfo
  {school} {State University of New York, Stony Brook (USA)} (\bibinfo {year}
  {1976})\BibitemShut {NoStop}%
\bibitem [{\citenamefont {Geesaman}\ \emph {et~al.}(1979)\citenamefont
  {Geesaman}, \citenamefont {McGrath}, \citenamefont {No\'e},\ and\
  \citenamefont {Malmin}}]{geesaman1979}%
  \BibitemOpen
  \bibfield  {author} {\bibinfo {author} {\bibfnamefont {D.~F.}\ \bibnamefont
  {Geesaman}}, \bibinfo {author} {\bibfnamefont {R.~L.}\ \bibnamefont
  {McGrath}}, \bibinfo {author} {\bibfnamefont {J.~W.}\ \bibnamefont {No\'e}},\
  and\ \bibinfo {author} {\bibfnamefont {R.~E.}\ \bibnamefont {Malmin}},\
  }\href {https://doi.org/10.1103/PhysRevC.19.1938} {\bibfield  {journal}
  {\bibinfo  {journal} {Phys. Rev. C}\ }\textbf {\bibinfo {volume} {19}},\
  \bibinfo {pages} {1938} (\bibinfo {year} {1979})}\BibitemShut {NoStop}%
\bibitem [{\citenamefont {Junde}(2009)}]{junde}%
  \BibitemOpen
  \bibfield  {author} {\bibinfo {author} {\bibfnamefont {H.}~\bibnamefont
  {Junde}},\ }\href {https://doi.org/https://doi.org/10.1016/j.nds.2009.10.001}
  {\bibfield  {journal} {\bibinfo  {journal} {Nucl. Data Sheets}\ }\textbf
  {\bibinfo {volume} {110}},\ \bibinfo {pages} {2689 } (\bibinfo {year}
  {2009})}\BibitemShut {NoStop}%
\bibitem [{\citenamefont {Dracoulis}\ and\ \citenamefont {Byrne}()}]{caesar}%
  \BibitemOpen
  \bibfield  {author} {\bibinfo {author} {\bibfnamefont {G.~D.}\ \bibnamefont
  {Dracoulis}}\ and\ \bibinfo {author} {\bibfnamefont {A.~P.}\ \bibnamefont
  {Byrne}},\ }\href@noop {} {\bibinfo {title} {Annual report {ANU-P}/1052
  (1989)}}\BibitemShut {NoStop}%
\bibitem [{sup()}]{supp}%
  \BibitemOpen
  \href@noop {} {}\bibinfo {note} {See Supplemental Material for details pertaining to the sum-event evaluation
  methods.}\BibitemShut {Stop}%
\bibitem [{\citenamefont {Palazzo}(2017)}]{palazzo}%
  \BibitemOpen
  \bibfield  {author} {\bibinfo {author} {\bibfnamefont {T.}~\bibnamefont
  {Palazzo}},\ }\emph {\bibinfo {title} {Spectroscopy and characterisation of
  high multipolarity transitions depopulating the metastable state in
  $^{53}$Fe}},\ \href {https://doi.org/10.25911/5d6510089d770} {Master's
  thesis},\ \bibinfo  {school} {The Australian National University, Canberra
  (Australia)} (\bibinfo {year} {2017})\BibitemShut {NoStop}%
\bibitem [{\citenamefont {Kib{\'e}di}\ \emph {et~al.}(2008)\citenamefont
  {Kib{\'e}di}, \citenamefont {Burrows}, \citenamefont {Trzhaskovskaya},
  \citenamefont {Davidson},\ and\ \citenamefont {Nestor~Jr.}}]{kibedi}%
  \BibitemOpen
  \bibfield  {author} {\bibinfo {author} {\bibfnamefont {T.}~\bibnamefont
  {Kib{\'e}di}}, \bibinfo {author} {\bibfnamefont {T.~W.}\ \bibnamefont
  {Burrows}}, \bibinfo {author} {\bibfnamefont {M.~B.}\ \bibnamefont
  {Trzhaskovskaya}}, \bibinfo {author} {\bibfnamefont {P.~M.}\ \bibnamefont
  {Davidson}},\ and\ \bibinfo {author} {\bibfnamefont {C.~W.}\ \bibnamefont
  {Nestor~Jr.}},\ }\href@noop {} {\bibfield  {journal} {\bibinfo  {journal}
  {Nucl. Instrum. Meth. A}\ }\textbf {\bibinfo {volume} {589}},\ \bibinfo
  {pages} {202} (\bibinfo {year} {2008})}\BibitemShut {NoStop}%
\bibitem [{\citenamefont {Band}\ \emph {et~al.}(2002)\citenamefont {Band},
  \citenamefont {Trzhaskovskaya}, \citenamefont {Nestor}, \citenamefont
  {Tikkanen},\ and\ \citenamefont {Raman}}]{band2002}%
  \BibitemOpen
  \bibfield  {author} {\bibinfo {author} {\bibfnamefont {I.}~\bibnamefont
  {Band}}, \bibinfo {author} {\bibfnamefont {M.}~\bibnamefont
  {Trzhaskovskaya}}, \bibinfo {author} {\bibfnamefont {C.}~\bibnamefont
  {Nestor}}, \bibinfo {author} {\bibfnamefont {P.}~\bibnamefont {Tikkanen}},\
  and\ \bibinfo {author} {\bibfnamefont {S.}~\bibnamefont {Raman}},\ }\href
  {https://doi.org/https://doi.org/10.1006/adnd.2002.0884} {\bibfield
  {journal} {\bibinfo  {journal} {At. Data Nucl. Data Tables}\ }\textbf
  {\bibinfo {volume} {81}},\ \bibinfo {pages} {1} (\bibinfo {year}
  {2002})}\BibitemShut {NoStop}%
\bibitem [{\citenamefont {Brown}\ and\ \citenamefont {Rae}(2014)}]{brown2014}%
  \BibitemOpen
  \bibfield  {author} {\bibinfo {author} {\bibfnamefont {B.}~\bibnamefont
  {Brown}}\ and\ \bibinfo {author} {\bibfnamefont {W.}~\bibnamefont {Rae}},\
  }\href {https://doi.org/https://doi.org/10.1016/j.nds.2014.07.022} {\bibfield
   {journal} {\bibinfo  {journal} {Nucl. Data Sheets}\ }\textbf {\bibinfo
  {volume} {120}},\ \bibinfo {pages} {115} (\bibinfo {year}
  {2014})}\BibitemShut {NoStop}%
\bibitem [{\citenamefont {Honma}\ \emph {et~al.}(2005)\citenamefont {Honma},
  \citenamefont {Otsuka}, \citenamefont {Brown},\ and\ \citenamefont
  {Mizusaki}}]{honma2005}%
  \BibitemOpen
  \bibfield  {author} {\bibinfo {author} {\bibfnamefont {M.}~\bibnamefont
  {Honma}}, \bibinfo {author} {\bibfnamefont {T.}~\bibnamefont {Otsuka}},
  \bibinfo {author} {\bibfnamefont {B.~A.}\ \bibnamefont {Brown}},\ and\
  \bibinfo {author} {\bibfnamefont {T.}~\bibnamefont {Mizusaki}},\ }\href
  {https://doi.org/10.1140/epjad/i2005-06-032-2} {\bibfield  {journal}
  {\bibinfo  {journal} {Euro Phys. J. A}\ }\textbf {\bibinfo {volume} {25}},\
  \bibinfo {pages} {499 } (\bibinfo {year} {2005})}\BibitemShut {NoStop}%
\bibitem [{\citenamefont {Poves}\ \emph {et~al.}(2001)\citenamefont {Poves},
  \citenamefont {Sánchez-Solano}, \citenamefont {Caurier},\ and\ \citenamefont
  {Nowacki}}]{poves2001}%
  \BibitemOpen
  \bibfield  {author} {\bibinfo {author} {\bibfnamefont {A.}~\bibnamefont
  {Poves}}, \bibinfo {author} {\bibfnamefont {J.}~\bibnamefont
  {Sánchez-Solano}}, \bibinfo {author} {\bibfnamefont {E.}~\bibnamefont
  {Caurier}},\ and\ \bibinfo {author} {\bibfnamefont {F.}~\bibnamefont
  {Nowacki}},\ }\href
  {https://doi.org/https://doi.org/10.1016/S0375-9474(01)00967-8} {\bibfield
  {journal} {\bibinfo  {journal} {Nucl. Phys. A}\ }\textbf {\bibinfo {volume}
  {694}},\ \bibinfo {pages} {157 } (\bibinfo {year} {2001})}\BibitemShut
  {NoStop}%
\bibitem [{\citenamefont {Brown}(1998)}]{brown1998}%
  \BibitemOpen
  \bibfield  {author} {\bibinfo {author} {\bibfnamefont {B.~A.}\ \bibnamefont
  {Brown}},\ }\href {https://doi.org/10.1103/PhysRevC.58.220} {\bibfield
  {journal} {\bibinfo  {journal} {Phys. Rev. C}\ }\textbf {\bibinfo {volume}
  {58}},\ \bibinfo {pages} {220} (\bibinfo {year} {1998})}\BibitemShut
  {NoStop}%
\bibitem [{\citenamefont {Brown}\ \emph {et~al.}(1983)\citenamefont {Brown},
  \citenamefont {Radhi},\ and\ \citenamefont {Wildenthal}}]{brown1983}%
  \BibitemOpen
  \bibfield  {author} {\bibinfo {author} {\bibfnamefont {B.}~\bibnamefont
  {Brown}}, \bibinfo {author} {\bibfnamefont {R.}~\bibnamefont {Radhi}},\ and\
  \bibinfo {author} {\bibfnamefont {B.}~\bibnamefont {Wildenthal}},\ }\href
  {https://doi.org/https://doi.org/10.1016/0370-1573(83)90001-7} {\bibfield
  {journal} {\bibinfo  {journal} {Phys. Rep.}\ }\textbf {\bibinfo {volume}
  {101}},\ \bibinfo {pages} {313 } (\bibinfo {year} {1983})}\BibitemShut
  {NoStop}%
\bibitem [{\citenamefont {Gloeckner}\ and\ \citenamefont
  {Lawson}(1975)}]{gloeckner1975}%
  \BibitemOpen
  \bibfield  {author} {\bibinfo {author} {\bibfnamefont {D.~H.}\ \bibnamefont
  {Gloeckner}}\ and\ \bibinfo {author} {\bibfnamefont {R.~D.}\ \bibnamefont
  {Lawson}},\ }\href {https://doi.org/10.1103/PhysRevC.11.1832} {\bibfield
  {journal} {\bibinfo  {journal} {Phys. Rev. C}\ }\textbf {\bibinfo {volume}
  {11}},\ \bibinfo {pages} {1832} (\bibinfo {year} {1975})}\BibitemShut
  {NoStop}%
\bibitem [{\citenamefont {Honma}\ \emph {et~al.}(2004)\citenamefont {Honma},
  \citenamefont {Otsuka}, \citenamefont {Brown},\ and\ \citenamefont
  {Mizusaki}}]{honma2004}%
  \BibitemOpen
  \bibfield  {author} {\bibinfo {author} {\bibfnamefont {M.}~\bibnamefont
  {Honma}}, \bibinfo {author} {\bibfnamefont {T.}~\bibnamefont {Otsuka}},
  \bibinfo {author} {\bibfnamefont {B.~A.}\ \bibnamefont {Brown}},\ and\
  \bibinfo {author} {\bibfnamefont {T.}~\bibnamefont {Mizusaki}},\ }\href
  {https://doi.org/10.1103/PhysRevC.69.034335} {\bibfield  {journal} {\bibinfo
  {journal} {Phys. Rev. C}\ }\textbf {\bibinfo {volume} {69}},\ \bibinfo
  {pages} {034335} (\bibinfo {year} {2004})}\BibitemShut {NoStop}%
\bibitem [{\citenamefont {du~Rietz}\ \emph {et~al.}(2004)\citenamefont
  {du~Rietz}, \citenamefont {Ekman}, \citenamefont {Rudolph}, \citenamefont
  {Fahlander}, \citenamefont {Dewald}, \citenamefont {M\"oller}, \citenamefont
  {Saha}, \citenamefont {Axiotis}, \citenamefont {Bentley}, \citenamefont
  {Chandler}, \citenamefont {de~Angelis}, \citenamefont {Della~Vedova},
  \citenamefont {Gadea}, \citenamefont {Hammond}, \citenamefont {Lenzi},
  \citenamefont {M\ifmmode~\u{a}\else \u{a}\fi{}rginean}, \citenamefont
  {Napoli}, \citenamefont {Nespolo}, \citenamefont {Rusu},\ and\ \citenamefont
  {Tonev}}]{durietz2004}%
  \BibitemOpen
  \bibfield  {author} {\bibinfo {author} {\bibfnamefont {R.}~\bibnamefont
  {du~Rietz}}, \bibinfo {author} {\bibfnamefont {J.}~\bibnamefont {Ekman}},
  \bibinfo {author} {\bibfnamefont {D.}~\bibnamefont {Rudolph}}, \bibinfo
  {author} {\bibfnamefont {C.}~\bibnamefont {Fahlander}}, \bibinfo {author}
  {\bibfnamefont {A.}~\bibnamefont {Dewald}}, \bibinfo {author} {\bibfnamefont
  {O.}~\bibnamefont {M\"oller}}, \bibinfo {author} {\bibfnamefont
  {B.}~\bibnamefont {Saha}}, \bibinfo {author} {\bibfnamefont {M.}~\bibnamefont
  {Axiotis}}, \bibinfo {author} {\bibfnamefont {M.~A.}\ \bibnamefont
  {Bentley}}, \bibinfo {author} {\bibfnamefont {C.}~\bibnamefont {Chandler}},
  \bibinfo {author} {\bibfnamefont {G.}~\bibnamefont {de~Angelis}}, \bibinfo
  {author} {\bibfnamefont {F.}~\bibnamefont {Della~Vedova}}, \bibinfo {author}
  {\bibfnamefont {A.}~\bibnamefont {Gadea}}, \bibinfo {author} {\bibfnamefont
  {G.}~\bibnamefont {Hammond}}, \bibinfo {author} {\bibfnamefont {S.~M.}\
  \bibnamefont {Lenzi}}, \bibinfo {author} {\bibfnamefont {N.}~\bibnamefont
  {M\ifmmode~\u{a}\else \u{a}\fi{}rginean}}, \bibinfo {author} {\bibfnamefont
  {D.~R.}\ \bibnamefont {Napoli}}, \bibinfo {author} {\bibfnamefont
  {M.}~\bibnamefont {Nespolo}}, \bibinfo {author} {\bibfnamefont
  {C.}~\bibnamefont {Rusu}},\ and\ \bibinfo {author} {\bibfnamefont
  {D.}~\bibnamefont {Tonev}},\ }\href
  {https://doi.org/10.1103/PhysRevLett.93.222501} {\bibfield  {journal}
  {\bibinfo  {journal} {Phys. Rev. Lett.}\ }\textbf {\bibinfo {volume} {93}},\
  \bibinfo {pages} {222501} (\bibinfo {year} {2004})}\BibitemShut {NoStop}%
\bibitem [{\citenamefont {Sagawa}(1979)}]{sagawa1979}%
  \BibitemOpen
  \bibfield  {author} {\bibinfo {author} {\bibfnamefont {H.}~\bibnamefont
  {Sagawa}},\ }\href {https://doi.org/10.1103/PhysRevC.19.506} {\bibfield
  {journal} {\bibinfo  {journal} {Phys. Rev. C}\ }\textbf {\bibinfo {volume}
  {19}},\ \bibinfo {pages} {506} (\bibinfo {year} {1979})}\BibitemShut
  {NoStop}%
\bibitem [{\citenamefont {Bohr}\ and\ \citenamefont {Mottelson}(1998)}]{BM}%
  \BibitemOpen
  \bibfield  {author} {\bibinfo {author} {\bibfnamefont {A.}~\bibnamefont
  {Bohr}}\ and\ \bibinfo {author} {\bibfnamefont {B.~R.}\ \bibnamefont
  {Mottelson}},\ }\href {https://doi.org/10.1142/3530} {\emph {\bibinfo {title}
  {Nuclear Structure}}}\ (\bibinfo  {publisher} {World Scientific Publishing
  Company},\ \bibinfo {year} {1998})\BibitemShut {NoStop}%
\bibitem [{\citenamefont {den Herder}\ \emph {et~al.}(1986)\citenamefont {den
  Herder}, \citenamefont {Hendriks}, \citenamefont {Jans}, \citenamefont
  {Keizer}, \citenamefont {Kramer}, \citenamefont {Lapik\'as}, \citenamefont
  {Quint}, \citenamefont {de~Witt~Huberts}, \citenamefont {Blok},\ and\
  \citenamefont {van~der Steenhoven}}]{denherder1986}%
  \BibitemOpen
  \bibfield  {author} {\bibinfo {author} {\bibfnamefont {J.~W.~A.}\
  \bibnamefont {den Herder}}, \bibinfo {author} {\bibfnamefont {J.~A.}\
  \bibnamefont {Hendriks}}, \bibinfo {author} {\bibfnamefont {E.}~\bibnamefont
  {Jans}}, \bibinfo {author} {\bibfnamefont {P.~H.~M.}\ \bibnamefont {Keizer}},
  \bibinfo {author} {\bibfnamefont {G.~J.}\ \bibnamefont {Kramer}}, \bibinfo
  {author} {\bibfnamefont {L.}~\bibnamefont {Lapik\'as}}, \bibinfo {author}
  {\bibfnamefont {E.~N.~M.}\ \bibnamefont {Quint}}, \bibinfo {author}
  {\bibfnamefont {P.~K.~A.}\ \bibnamefont {de~Witt~Huberts}}, \bibinfo {author}
  {\bibfnamefont {H.~P.}\ \bibnamefont {Blok}},\ and\ \bibinfo {author}
  {\bibfnamefont {G.}~\bibnamefont {van~der Steenhoven}},\ }\href
  {https://doi.org/10.1103/PhysRevLett.57.1843} {\bibfield  {journal} {\bibinfo
   {journal} {Phys. Rev. Lett.}\ }\textbf {\bibinfo {volume} {57}},\ \bibinfo
  {pages} {1843} (\bibinfo {year} {1986})}\BibitemShut {NoStop}%
\bibitem [{\citenamefont {M\"uther}\ \emph {et~al.}(1995)\citenamefont
  {M\"uther}, \citenamefont {Polls},\ and\ \citenamefont
  {Dickhoff}}]{muther1995}%
  \BibitemOpen
  \bibfield  {author} {\bibinfo {author} {\bibfnamefont {H.}~\bibnamefont
  {M\"uther}}, \bibinfo {author} {\bibfnamefont {A.}~\bibnamefont {Polls}},\
  and\ \bibinfo {author} {\bibfnamefont {W.~H.}\ \bibnamefont {Dickhoff}},\
  }\href {https://doi.org/10.1103/PhysRevC.51.3040} {\bibfield  {journal}
  {\bibinfo  {journal} {Phys. Rev. C}\ }\textbf {\bibinfo {volume} {51}},\
  \bibinfo {pages} {3040} (\bibinfo {year} {1995})}\BibitemShut {NoStop}%
\bibitem [{\citenamefont {Dickhoff}(2010)}]{dickhoff2010}%
  \BibitemOpen
  \bibfield  {author} {\bibinfo {author} {\bibfnamefont {W.~H.}\ \bibnamefont
  {Dickhoff}},\ }\href {https://doi.org/10.1088/0954-3899/37/6/064007}
  {\bibfield  {journal} {\bibinfo  {journal} {J. Phys. G}\ }\textbf {\bibinfo
  {volume} {37}},\ \bibinfo {pages} {064007} (\bibinfo {year}
  {2010})}\BibitemShut {NoStop}%
\bibitem [{\citenamefont {Barbieri}(2009)}]{barbieri2009}%
  \BibitemOpen
  \bibfield  {author} {\bibinfo {author} {\bibfnamefont {C.}~\bibnamefont
  {Barbieri}},\ }\href {https://doi.org/10.1103/PhysRevLett.103.202502}
  {\bibfield  {journal} {\bibinfo  {journal} {Phys. Rev. Lett.}\ }\textbf
  {\bibinfo {volume} {103}},\ \bibinfo {pages} {202502} (\bibinfo {year}
  {2009})}\BibitemShut {NoStop}%
\bibitem [{\citenamefont {Lapikás}(1993)}]{lapikas1993}%
  \BibitemOpen
  \bibfield  {author} {\bibinfo {author} {\bibfnamefont {L.}~\bibnamefont
  {Lapikás}},\ }\href
  {https://doi.org/https://doi.org/10.1016/0375-9474(93)90630-G} {\bibfield
  {journal} {\bibinfo  {journal} {Nucl. Phys. A}\ }\textbf {\bibinfo {volume}
  {553}},\ \bibinfo {pages} {297} (\bibinfo {year} {1993})}\BibitemShut
  {NoStop}%
\bibitem [{\citenamefont {Kramer}\ \emph {et~al.}(2001)\citenamefont {Kramer},
  \citenamefont {Blok},\ and\ \citenamefont {Lapikás}}]{kramer2001}%
  \BibitemOpen
  \bibfield  {author} {\bibinfo {author} {\bibfnamefont {G.}~\bibnamefont
  {Kramer}}, \bibinfo {author} {\bibfnamefont {H.}~\bibnamefont {Blok}},\ and\
  \bibinfo {author} {\bibfnamefont {L.}~\bibnamefont {Lapikás}},\ }\href
  {https://doi.org/https://doi.org/10.1016/S0375-9474(00)00379-1} {\bibfield
  {journal} {\bibinfo  {journal} {Nucl. Phys. A}\ }\textbf {\bibinfo {volume}
  {679}},\ \bibinfo {pages} {267} (\bibinfo {year} {2001})}\BibitemShut
  {NoStop}%
\bibitem [{\citenamefont {Kay}\ \emph {et~al.}(2013)\citenamefont {Kay},
  \citenamefont {Schiffer},\ and\ \citenamefont {Freeman}}]{kay2013}%
  \BibitemOpen
  \bibfield  {author} {\bibinfo {author} {\bibfnamefont {B.~P.}\ \bibnamefont
  {Kay}}, \bibinfo {author} {\bibfnamefont {J.~P.}\ \bibnamefont {Schiffer}},\
  and\ \bibinfo {author} {\bibfnamefont {S.~J.}\ \bibnamefont {Freeman}},\
  }\href {https://doi.org/10.1103/PhysRevLett.111.042502} {\bibfield  {journal}
  {\bibinfo  {journal} {Phys. Rev. Lett.}\ }\textbf {\bibinfo {volume} {111}},\
  \bibinfo {pages} {042502} (\bibinfo {year} {2013})}\BibitemShut {NoStop}%
\bibitem [{\citenamefont {Berggren}(1965)}]{berggren1965}%
  \BibitemOpen
  \bibfield  {author} {\bibinfo {author} {\bibfnamefont {T.}~\bibnamefont
  {Berggren}},\ }\href
  {https://doi.org/https://doi.org/10.1016/0029-5582(65)90440-2} {\bibfield
  {journal} {\bibinfo  {journal} {Nucl. Phys.}\ }\textbf {\bibinfo {volume}
  {72}},\ \bibinfo {pages} {337} (\bibinfo {year} {1965})}\BibitemShut
  {NoStop}%
\bibitem [{\citenamefont {Caprio}(2005)}]{caprio2005}%
  \BibitemOpen
  \bibfield  {author} {\bibinfo {author} {\bibfnamefont {M.}~\bibnamefont
  {Caprio}},\ }\href
  {https://doi.org/https://doi.org/10.1016/j.cpc.2005.04.010} {\bibfield
  {journal} {\bibinfo  {journal} {Comput. Phys. Commun.}\ }\textbf {\bibinfo
  {volume} {171}},\ \bibinfo {pages} {107} (\bibinfo {year}
  {2005})}\BibitemShut {NoStop}%
\end{thebibliography}

\begin{thebibliography}{4}%
\makeatletter
\providecommand \@ifxundefined [1]{%
 \@ifx{#1\undefined}
}%
\providecommand \@ifnum [1]{%
 \ifnum #1\expandafter \@firstoftwo
 \else \expandafter \@secondoftwo
 \fi
}%
\providecommand \@ifx [1]{%
 \ifx #1\expandafter \@firstoftwo
 \else \expandafter \@secondoftwo
 \fi
}%
\providecommand \natexlab [1]{#1}%
\providecommand \enquote  [1]{``#1''}%
\providecommand \bibnamefont  [1]{#1}%
\providecommand \bibfnamefont [1]{#1}%
\providecommand \citenamefont [1]{#1}%
\providecommand \href@noop [0]{\@secondoftwo}%
\providecommand \href [0]{\begingroup \@sanitize@url \@href}%
\providecommand \@href[1]{\@@startlink{#1}\@@href}%
\providecommand \@@href[1]{\endgroup#1\@@endlink}%
\providecommand \@sanitize@url [0]{\catcode `\\12\catcode `\$12\catcode
  `\&12\catcode `\#12\catcode `\^12\catcode `\_12\catcode `\%12\relax}%
\providecommand \@@startlink[1]{}%
\providecommand \@@endlink[0]{}%
\providecommand \url  [0]{\begingroup\@sanitize@url \@url }%
\providecommand \@url [1]{\endgroup\@href {#1}{\urlprefix }}%
\providecommand \urlprefix  [0]{URL }%
\providecommand \Eprint [0]{\href }%
\providecommand \doibase [0]{http://dx.doi.org/}%
\providecommand \selectlanguage [0]{\@gobble}%
\providecommand \bibinfo  [0]{\@secondoftwo}%
\providecommand \bibfield  [0]{\@secondoftwo}%
\providecommand \translation [1]{[#1]}%
\providecommand \BibitemOpen [0]{}%
\providecommand \bibitemStop [0]{}%
\providecommand \bibitemNoStop [0]{.\EOS\space}%
\providecommand \EOS [0]{\spacefactor3000\relax}%
\providecommand \BibitemShut  [1]{\csname bibitem#1\endcsname}%
\let\auto@bib@innerbib\@empty
\bibitem [{pap()}]{paper}%
  \BibitemOpen
  \href@noop {} {}\bibinfo {note} {T.~Palazzo $et~al$., Phys. Rev. Lett.
  (Current manuscript).}\BibitemShut {Stop}%
\bibitem [{\citenamefont {Honma}\ \emph {et~al.}(2005)\citenamefont {Honma},
  \citenamefont {Otsuka}, \citenamefont {Brown},\ and\ \citenamefont
  {Mizusaki}}]{honma2005}%
  \BibitemOpen
  \bibfield  {author} {\bibinfo {author} {\bibfnamefont {M.}~\bibnamefont
  {Honma}}, \bibinfo {author} {\bibfnamefont {T.}~\bibnamefont {Otsuka}},
  \bibinfo {author} {\bibfnamefont {B.~A.}\ \bibnamefont {Brown}}, \ and\
  \bibinfo {author} {\bibfnamefont {T.}~\bibnamefont {Mizusaki}},\ }\href
  {\doibase 10.1140/epjad/i2005-06-032-2} {\bibfield  {journal} {\bibinfo
  {journal} {Euro Phys. J. A}\ }\textbf {\bibinfo {volume} {25}},\ \bibinfo
  {pages} {499 } (\bibinfo {year} {2005})}\BibitemShut {NoStop}%
\bibitem [{\citenamefont {Poves}\ \emph {et~al.}(2001)\citenamefont {Poves},
  \citenamefont {Sánchez-Solano}, \citenamefont {Caurier},\ and\ \citenamefont
  {Nowacki}}]{poves2001}%
  \BibitemOpen
  \bibfield  {author} {\bibinfo {author} {\bibfnamefont {A.}~\bibnamefont
  {Poves}}, \bibinfo {author} {\bibfnamefont {J.}~\bibnamefont
  {Sánchez-Solano}}, \bibinfo {author} {\bibfnamefont {E.}~\bibnamefont
  {Caurier}}, \ and\ \bibinfo {author} {\bibfnamefont {F.}~\bibnamefont
  {Nowacki}},\ }\href {\doibase https://doi.org/10.1016/S0375-9474(01)00967-8}
  {\bibfield  {journal} {\bibinfo  {journal} {Nucl. Phys. A}\ }\textbf
  {\bibinfo {volume} {694}},\ \bibinfo {pages} {157 } (\bibinfo {year}
  {2001})}\BibitemShut {NoStop}%
\bibitem [{\citenamefont {Gloeckner}\ and\ \citenamefont
  {Lawson}(1975)}]{gloeckner1975}%
  \BibitemOpen
  \bibfield  {author} {\bibinfo {author} {\bibfnamefont {D.~H.}\ \bibnamefont
  {Gloeckner}}\ and\ \bibinfo {author} {\bibfnamefont {R.~D.}\ \bibnamefont
  {Lawson}},\ }\href {\doibase 10.1103/PhysRevC.11.1832} {\bibfield  {journal}
  {\bibinfo  {journal} {Phys. Rev. C}\ }\textbf {\bibinfo {volume} {11}},\
  \bibinfo {pages} {1832} (\bibinfo {year} {1975})}\BibitemShut {NoStop}%
\end{thebibliography}

%

\clearpage

\textbf{Supplemental Material}: Reference \cite{paper} presents the first experimental confirmation of hexacontatetrapole, $E6$ $\gamma$ decay to include a detailed consideration of sum contributions to the measured $\gamma$-ray yields. The different approaches used to evaluate the sum contributions are described below with results summarised in Table \ref{stab1}. The sum-component fractions of the total 3041-keV $\gamma$-ray yield determined by each method were 48(11)$\%$ ($Experimental$), 47(25)$\%$ ($Geometric$), 46(6)$\%$ ($Computational$) and 42(11)$\%$ ($Monte~Carlo$), respectively. 


\begin{figure}[b!]
\includegraphics[width=8.5 cm]{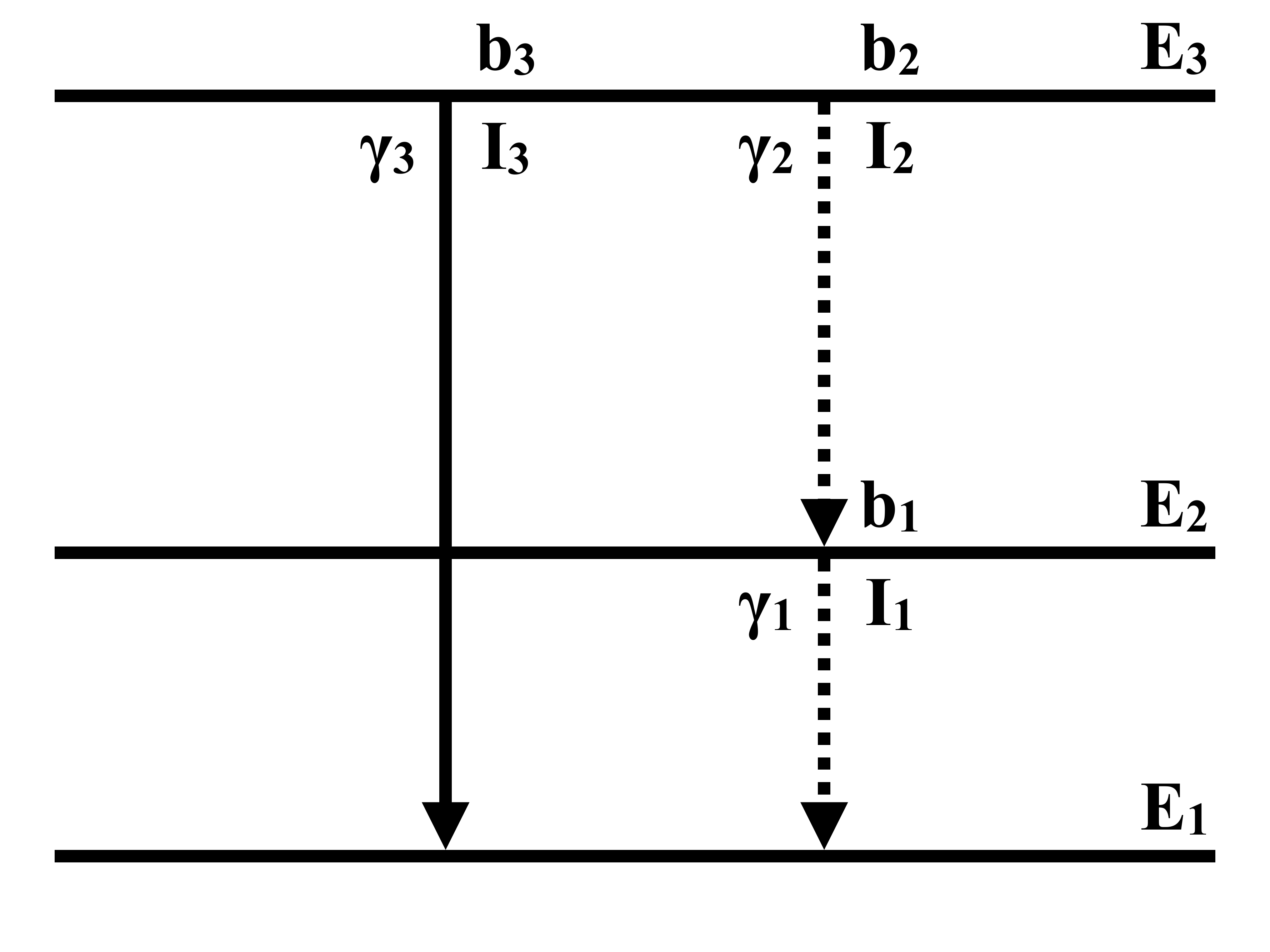}
\caption{\label{sfig1} Example level scheme and $\gamma$-ray transitions used to explain the methods of determining sum contributions of measured $\gamma$-ray yields described in the text. An excited state, $E_3$, has two relaxation pathways: one direct to $E_1$, and the other through a cascade of $\gamma$ rays via an intermediate level, $E_2$. Each transition yield is characterised by its intensity ($I$), branching ratio ($b$) and $\gamma$-ray detection efficiency ($\varepsilon$). }
\end{figure}

$Experimental$: In the example shown in Fig.~\ref{sfig1}, an excited state, E$_3$, relaxes via two distinct pathways: a single $\gamma$ decay ($\gamma_3$) direct to E$_1$ and a two-photon cascade ($\gamma_2$ and $\gamma_1$) via E$_2$. The relative probability of each pathway taking place is given by the branching ratios of the transitions that depopulate E$_3$ (b$_3$ and b$_2$). The measured yield of $\gamma_3$ ($Y_3$) includes the number of $\gamma_3$ decays (${I}_{3})$, corrected for detection efficiency ($\varepsilon_3$), and the additional sum component from $\gamma_2$ and $\gamma_1$ ($S_{2,1}$), such that: \\[-0.9 cm]

\begin{equation}
{Y}_{3}~=~{I}_{3} \cdot \varepsilon_3 + S_{2,1}, 
\label{seq1}
\end{equation}

\noindent
where: \\[-0.9 cm]

\begin{equation}
{S}_{2,1}~=~{I}_{2} \cdot \varepsilon_2 \cdot b_1 \cdot \varepsilon_1 \cdot \overline{W}_{2,1}(0), 
\label{seq2}
\end{equation}

\noindent and $\overline{W}_{2,1}(0)$ is the angular correlation of $\gamma_2$ and $\gamma_1$ at $\approx$ 0$^{\circ}$ averaged over the solid angle subtended by the detector. If more than one cascade pathway exists, then a sum over all combinations of sum contributions must be considered. 

The yield of the 2029-keV full-energy sum peak observed from the $\gamma$-decay of \nuc{53m}{Fe}, which can $only$ occur though summing of the 701-keV and 1328-keV $\gamma$-rays, can be directly measured and scaled to estimate the sum-coincidence components of the other transitions. Using the notation of Equation~\eqref{seq2}: \\[-0.9 cm]

\begin{multline}
{Y}_{2029}~=~{S}_{2029}~=~{I}_{701} \cdot \varepsilon_{701} \cdot b_{1011} \cdot b_{1328} \cdot \varepsilon_{1328} \\ 
\times {\overline{W}_{701,1328}(0)}. 
\label{seq3}
\end{multline}

\noindent
Expressions that connect the two- and three-fold sum components of each transition to ${S}_{2029}$, other measured $\gamma$-ray yields and branching ratios, detection efficiencies and calculated angular correlations between pairs of $\gamma$ rays can then be deduced. For example, the contributions to the 3041-keV full-energy peak are given by:  \\[-0.9 cm]

\begin{multline} 
{S}_{701,2338}~=~{S}_{2029} \cdot \left( \frac{Y_{2338}-Y_{1011} \cdot \varepsilon_{1328} \cdot {\overline{W}_{1011,2338}(0)}}{Y_{1011}} \right)\\
\times \left(\frac{\varepsilon_{1011}}{\varepsilon_{1328}} \right) \cdot \left(\frac{\overline{W}_{701,2338}(0)}{\overline{W}_{701,1328}(0)} \right), 
\label{seq4}
\end{multline}

\begin{multline} 
{S}_{1713,1328}~=~{S}_{2029} \cdot \left( \frac{Y_{1713}-Y_{1011} \cdot \varepsilon_{701} \cdot {\overline{W}_{701,1011}(0)}}{Y_{1011}} \right)\\
\times \left(\frac{\varepsilon_{1011}}{\varepsilon_{701}} \right) \cdot \left(\frac{\overline{W}_{1713,1328}(0)}{\overline{W}_{701,1328}(0)} \right),
\label{seq5}
\end{multline}

\begin{equation}
{S}_{701,1011,1328}~=~{S}_{2029} \cdot \varepsilon_{1011} \cdot \left(\frac{\overline{W}_{701,1011,1328}(0)}{\overline{W}_{701,1328}(0)} \right).
\label{seq6}
\end{equation}

\noindent
Similar expressions can be defined for the sum components of the 1713-keV and 2338-keV transitions; these corrections are $\approx 10\%$ and $\approx 1\%$, respectively. 
%

$Geometric$: The sum contributions can be directly inferred by considering the change in counting efficiency between different detector geometries. The three $\gamma$-ray detectors mounted on adjustable rails were moved radially outwards by $\approx$ 3.5~cm to reduce their solid angle coverage. This decreased the expected full-energy peak yields by a reduction factor, $r_{\varepsilon}$, for a `real', single $\gamma$-ray event, $r_{\varepsilon}^2$ for a sum-coincidence event and $r_{\varepsilon}^3$ for a triple-sum-coincidence event. The measured total yields for each $\gamma$~ray, in both the `near' and `far' geometries, can be reduced to a single expression connecting the real component to the total measured yields and respective detection efficiencies. In this work, $r_{\varepsilon}$~=~2.01(6) was deduced from measurement of the absolute detection efficiency for both geometries. Gamma-ray intensities determined from this geometric approach were consistent with the Experimental method, giving further confidence in the results. However, these suffered from larger experimental uncertainties and are not included in the final branching-ratio analysis. 
%

$Computational$: Following Equation \eqref{seq1}, a general expression for $Y_{3041}$ can be defined as follows: \\[-0.9 cm]

\begin{align} 
{Y}_{3041}~=~& I_{3041} \cdot \varepsilon_{3041}  \label{seq7} \\ \nonumber 
    			  & + I_{701} \cdot b_{2338} \cdot \varepsilon_{701} \cdot \varepsilon_{2338} \cdot \overline{W}_{701,2338}(\theta) \\ \nonumber
    			  & + I_{1713} \cdot b_{1328} \cdot \varepsilon_{1713} \cdot \varepsilon_{1328} \cdot \overline{W}_{1713,1328}(\theta) \\ \nonumber
    			  & + I_{701} \cdot b_{1011} \cdot b_{1328} \cdot \varepsilon_{701} \cdot \varepsilon_{1011} \cdot \varepsilon_{1328} \\ \nonumber
			  &	 \hspace{2.5cm}  \times \overline{W}_{701,1011,1328}(\theta).
\end{align}

Since $Y^{real}_i = I_i \cdot \varepsilon_i$ for non-sum events and $Y_1 = I_2 \cdot b_2 \cdot \varepsilon_1 $ for sequential cascades of $\gamma$ rays (such as the $\gamma_2$ to $\gamma_1$ cascade in Fig.~\ref{sfig1}), Equation \eqref{seq7} can be further reduced to a single expression that only includes quantities that were measured directly in the experiment---$\gamma$-ray yields and detection efficiencies---and calculated angular correlation coefficients, such that: \\[-0.9 cm]

\begin{equation}
{Y}_{3041}~=~Y^{real}_{3041} + a + b + c,  \\[-0.5 cm]
\label{seq8}
\end{equation}

\noindent
where: \\[-1 cm]

\begin{align}
a = (Y_{2338} & - Y_{1011} \cdot \varepsilon_{1328} \cdot \overline{W}_{1011,1328}(\theta)) \cdot \varepsilon_{701} \label{seq9} \\ \nonumber 
	& \times \overline{W}_{701,2338}(\theta), \\ \nonumber
b = (Y_{1713} & - Y_{1011} \cdot \varepsilon_{701} \cdot \overline{W}_{701,1011}(\theta))\cdot \varepsilon_{1328} \\ \nonumber
& \times \overline{W}_{1713,1328}(\theta), \\ \nonumber
c = (Y_{1328} & \cdot \varepsilon_{701} \cdot \varepsilon_{1011} \\ \nonumber
&+ (Y_{1011} \cdot \varepsilon_{701} \cdot \overline{W}_{701,1011}(\theta) - Y_{1713}) \\ \nonumber
& \times \frac{\varepsilon_{701} \cdot \varepsilon_{1011} \cdot \varepsilon_{1328}}{\varepsilon_{1713}}) \cdot \overline{W}_{701,1011,1328}(\theta),
\end{align}

\noindent
and $S_{3041} = a + b + c$. Following this methodology, similar expressions can be determined for the sum contributions to the other $\gamma$ rays and the 2029-keV sum peak: \\[-0.9cm]

\begin{align}
{Y}_{1713}~=~& I_{1713} \cdot \varepsilon_{1713} \label{seq10} \\ \nonumber 
    			  & + I_{701} \cdot b_{1011} \cdot \varepsilon_{701} \cdot \varepsilon_{1011} \cdot \overline{W}_{701,1011}(\theta), \\ \nonumber
		~=~ & Y^{real}_{1713} + (Y_{1011} \cdot \varepsilon_{701} \cdot \overline{W}_{701,1011}(\theta)).
\end{align}

\begin{align}
{Y}_{2338}~=~& I_{2338} \cdot \varepsilon_{2338} \label{seq11} \\ \nonumber 
    			  & + I_{1011} \cdot b_{1328} \cdot \varepsilon_{1011} \cdot \varepsilon_{1328} \cdot \overline{W}_{1011,1328}(\theta), \\ \nonumber
		~=~ & Y^{real}_{2338} + (Y_{1011} \cdot \varepsilon_{1328} \cdot \overline{W}_{1011,1328}(\theta)).
\end{align}

\begin{align}
{Y}_{2029}~=~ & I_{701} \cdot b_{1011}  \cdot b_{1328} \cdot \varepsilon_{701} \cdot \varepsilon_{1328} \cdot \overline{W}_{1701,1328}(\theta), \label{seq12} \\ \nonumber 
		~=~ & Y_{1328} \cdot \varepsilon_{701} \cdot \overline{W}_{1701,1328}(\theta).
\end{align}

Equations \eqref{seq8}, \eqref{seq9}, \eqref{seq10} \eqref{seq11}, \eqref{seq12} now define the sum contributions to each peak in terms of the experimentally observed peak yields that are inclusive of both the real and sum components. These observed peak yields and their uncertainties can then be used to determine the summing contributions. Uncertainties in the sum components were evaluated using a Monte Carlo methodology. The $\gamma$-ray yields were randomly sampled using Gaussian distributions centred on the measured values with widths defined by their uncertainties. This resulted in distributions of real and sum-coincidence yields, from which the mean and standard deviation were used in the subsequent analysis. 

$Monte~Carlo$: From the experimental and computational methods, a set of branching ratios were deduced from the corrected $\gamma$-ray yields, removing the sum components. Consistency of these results was confirmed by performing a Monte Carlo simulation of the $\gamma$ decay of $^{53m}$Fe and evaluating the implied summing contributions as an additional check. 

The Monte Carlo simulation was developed to model the $\gamma$ decay of \nuc{53m}{Fe} and evaluate the associated summing contributions. In the model, decay of \nuc{53m}{Fe} proceeds via randomised pathways that are weighted by the measured transition branching ratios of this work. The simulation considers each individual detector efficiency and takes account of angular-correlation effects. A sum event is recorded when two or more $\gamma$ rays from a cascade are recorded in the same detector. The number of simulated decays was fixed as the number of \nuc{53m}{Fe} decay events that occurred in the experiments for the near ($\sim$208M) and far ($\sim$214M) geometries, respectively. Separate simulations were used to estimate statistical errors: five iterations of 10 million \nuc{53m}{Fe} decay events; one iteration of 100 million events, and two iterations of one billion events. Results of the MC simulation for the total yields and sum components are consistent with the experimental value for the 3041-keV transition. MC-simulated yields of the other real $\gamma$ rays are within 5$\%$ of the corresponding experimental values. 

Results obtained from the different approaches are summarised in Table \ref{stab1}. Importantly, evaluations of the sum contributions to each of the $\gamma$ rays agree between the experimental and computational methods, and they are also in agreement with the Monte Carlo predictions. 

\begin{table*}[th!]
\centering
\caption{\label{stab1}Summary of the raw data for each full-energy peak area observed in the decay of \nuc{53m}{Fe} in both the `near' and `far' detector geometries discussed in the text of Ref.~\cite{paper}. Total yields of each full-energy peak ($Y_{\gamma}^{}$) are shown for the experimental (Expt.) and Monte Carlo (MC) methods; these include any possible two- and three-fold summing events. Sum contributions ($S_{i}^{}$) to relevant full-energy peak yields are shown for experimental (Expt.), computational (Comp.) and Monte Carlo (MC) methods. All three methods give consistent results for each $\gamma$-ray energy, including the different sum contributions observed in the `Near' and `Far' geometries.}
\vspace{0.1cm}
\label{table1}
\begin{tabular*}{18cm}{@{\extracolsep{\fill}}lcccccccccc}
\hline
\hline  \\[-0.3cm]
	&	\multicolumn{5}{c}{Near}	&\multicolumn{5}{c}{Far}	\\[0.1cm]
\cline{1-1} \cline{2-6}  \cline{7-11} \\[-0.3cm]
 $E_{\gamma}$	     &\multicolumn{2}{c}{Total yield, $Y_{\gamma}^{}$}	& \multicolumn{3}{c}{Sum, $S_{i}^{}$}	& \multicolumn{2}{c}{Total yield, $Y_{\gamma}^{}$}	& \multicolumn{3}{c}{Sum, $S_{i}^{}$}		\\[0.1cm]
\cline{1-1} \cline{2-3}  \cline{4-6}  \cline{7-8}  \cline{9-11} \\[-0.3cm]
(keV)	 &Expt. 			& MC 	 		& Expt.			& Comp.	& MC 		&Expt. 			& MC 			& Expt.		& Comp.	& MC	\\[0.1cm]
\hline 
701 		& 942815(2855) 	& 942731(4317) 	& -				& -		& -			& 691684(2449)	& 690508(3162) 	& -			&- 		& - 		\\
1011 	& 585573(3110) 	& 578761(2412) 	& -				& -		& -			& 436114(2184)	& 424867(1770)	&  -			& -		& -		\\
1328 	& 507607(2192)  	& 485255(2367) 	& -				& -		& -			& 378218(1850)	& 356993(1741)	&  -			& -		& - 		\\
1713 	& 6190(263)      	& 6136(269)    		& 794(189)		& 730(27)	& 732(106)	& 4427(222)		& 4298(188)		& 500(155)	& 375(11)	& 366(53) \\
2029 	& 614(140)   		& 542(104)	    	& 614(140)		& 630(23)	& 542(104)	& 387(119) 		& 310(60)			& 387(119)	& 329(9) 	& 310(60) \\
2338 	& 88164(861)	 	& 88271(1589)   	& 762(181)		& 760(28)	& 720(95) 		& 68356(768) 		&	65264(1175)   	& 480(149)	& 390(11)	& 366(48) \\
3041 	& 394(62)			& 338(86)  		& 178(42)   		& 185(7)	& 163(46)		& 217(48) 			& 212(54) 		 	& 118(37)		& 99(3)	& 82(23) \\
\hline
\hline
\end{tabular*}
\end{table*}

Table \ref{stab2} shows the complete set of calculated matrix elements, as described in the text of Ref.~\cite{paper}. Theoretical values of proton and neutron components ($\mathcal{A}_{p,n}$) of the $E6$, $M5$ and $E4$ matrix elements are provided. 

\begin{table*}[ht!]
\centering
\caption{\label{stab2}Theoretical values of proton and neutron components ($\mathcal{A}_{p,n}$) of $E6$, $M5$ and $E4$ matrix elements discussed in the text of Ref.~\cite{paper}. These values were calculated with the Skx radial wavefunctions and an oscillator length of $ b= 1.937 $~fm. For the electric transitions, $E \lambda$, $\mathcal{A}_p$ (e fm$^{\lambda}$) is obtained with $e_p=1.0$ and $e_n=0.0$, while $\mathcal{A}_n$ is calculated with $e_p=0.0$ and $e_n=1.0$. For the $M5$ transition, $\mathcal{A}_{p,n}$ ($\mu_N$ fm$^{\lambda -1}$) is calculated with the free-nucleon $g$ factors, and the matrix element $\mathcal{M}~=~(\mathcal{A}_p~+~\mathcal{A}_n)$. Uncertainties in the calculated matrix-element components are $\pm$(18,15,12)\% for $ L=(6,5,4) $, respectively. Experimental matrix elements ($\mathcal{M}_p^{\rm{expt.}}$) are determined from the $B(\sigma L$) values measured in this work.}
\vspace{0.1cm}
\label{table3}
\begin{tabular*}{18cm}{@{\extracolsep{\fill}}ccclll}
\hline 
\hline \\[-0.3cm]
$\sigma L$ 	&	Model Space &	Hamiltonian	 &	$\mathcal{A}_p$$\times$10$^3$	&	$\mathcal{A}_n$$\times$10$^3$			&	$\mathcal{M}$$\times$10$^3$		  \\[0.05cm]
\hline \\[-0.3cm]
		 	&			     &			 &	\multicolumn{3}{c}{(e fm$^{\lambda}$, $\mu_N$ fm$^{\lambda -1}$)	}								  \\[0.05cm]
\hline 
\hline  \\[-0.3cm]
		
$E6$ 	& $(fp)^{13}$	& GFPX1A \cite{honma2005}		&	3.34(60)		&	0.14(3)		& 		-		 \\
		&	& KB3G \cite{poves2001}	& 	3.69(66)		&	0.29(5)	& 	-			  			 \\
		&  	& Average		& 	3.52(63)	&	0.22(4)	& 				-  			 \\
		& ($f_{7/2})^{13}$	& 	\cite{gloeckner1975}	&	5.73(103)	&	0.15(3)		& 	-			  		  \\[0.1cm] \cline{5-6} \\[-0.3cm]
				 			\multicolumn{6}{r}{$\mathcal{M}_p^{\rm{expt.}}\times10^3$: 2.29(35)	 e fm$^{6}$} 	 \\[0.05cm]
\hline  \\[-0.3cm]

$M5$	& $(fp)^{13}$	& GFPX1A			&	4.71(71)	&	-0.16(2) 	& 	4.55(71)  			\\
		&	& KB3G			&	5.47(82)		&	-0.06(1) 	& 	5.41(82)						\\
		&  	& Average		& 	5.09(76)	&	-0.11(2)		& 	4.98(76)				  			 \\
		& ($f_{7/2})^{13}$	& \cite{gloeckner1975}	&	8.40(122)	&	0.14(2)		& 	8.54(126)				  		 \\[0.1cm] \cline{5-6} \\[-0.3cm]
 			\multicolumn{6}{r}{$\mathcal{M}_p^{\rm{expt.}}\times10^3$: 2.57(6) $\mu_N$ fm$^{4}$} 	 \\[0.05cm]

\hline  \\[-0.3cm]

$E4$	 	& $(fp)^{13}$	& GFPX1A			&	0.129(15)	& 	0.039(5)	&	-		 \\
		&	& KB3G			&	0.155(19)	& 	0.051(6)	&		-			  \\
		&  	& Average		& 	0.142(17)		&	0.045(5)		& 	-			  			 \\
		&($f_{7/2})^{13}$	& \cite{gloeckner1975}	&	0.216(26)	& 	0.019(2)	&	-					   \\[0.1cm] \cline{5-6} \\[-0.3cm]
			 			\multicolumn{6}{r}{$\mathcal{M}_p^{\rm{expt.}}\times10^3$: 0.1137(5) e fm$^{4}$} 	 \\[0.05cm]

\hline
\hline  \\[-0.3cm]

\end{tabular*}
\end{table*}

\linespread{1}

%


\end{document}